\documentclass[prl,twocolumn,superscriptaddress,floatfix,showpacs]{revtex4}
\usepackage{graphicx,amsfonts,amssymb,amsmath, hyperref}

\newif\ifhyper
\hypertrue
\ifhyper
\hypersetup{
   citecolor = {green},
   colorlinks = {true}, 
   urlcolor = {blue} 
}
\fi

\newcommand{\beq}{\begin{equation}}
\newcommand{\eeq}{\end{equation}}
\newcommand{\beqa}{\begin{eqnarray}}
\newcommand{\eeqa}{\end{eqnarray}}
\newcommand{\ket} [1] {\vert #1 \rangle}

\newcommand{\braket}[2]{\langle #1 | #2 \rangle}

\def\ket#1{\vert#1\rangle}

\def\Longarrow{\protect\@lra}
\def\@lra{\relbar\joinrel\relbar\joinrel\relbar\joinrel%
          \relbar\joinrel\rightarrow}

\begin{document}

\title{Bounds on universal quantum computation with perturbed 2d cluster states}

\author{Rom\'an Or\'us}
\affiliation{Institute of Physics, Johannes Gutenberg University, 55099 Mainz, Germany}
\affiliation{Max-Planck-Institut f\"ur Quantenoptik, Hans-Kopfermann-Str. 1, 85748 Garching, Germany}

\author{Henning Kalis}
\affiliation{Albert-Ludwigs-Universit\"at Freiburg, Physikalisches Institut, Hermann-Herder-Strasse 3, 79104 Freiburg, Germany}
\affiliation{Lehrstuhl f\"ur Theoretische Physik I, Otto-Hahn-Str. 4, TU Dortmund, D-44221 Dortmund, Germany}

\author{Marcel Bornemann}

\author{Kai Phillip Schmidt}
\affiliation{Lehrstuhl f\"ur Theoretische Physik I, Otto-Hahn-Str. 4, TU Dortmund, D-44221 Dortmund, Germany}

\begin{abstract}

Motivated by the possibility of universal quantum computation under noise perturbations, we compute the phase diagram of the 2d cluster state Hamiltonian in the presence of Ising terms and magnetic fields. Unlike in previous analysis of perturbed 2d cluster states, we find strong evidence of a very well defined cluster phase, separated from a polarized phase by a line of 1st and 2nd order transitions compatible with the 3d Ising universality class and a tricritical end point. The phase boundary sets an upper bound for the amount of perturbation in the system so that its ground state is still useful for measurement-based quantum computation purposes. Moreover, we also compute the local fidelity with the unperturbed 2d cluster state. Besides a classical approximation, we determine the phase diagram by combining series expansions and variational infinite Projected entangled-Pair States (iPEPS) methods. Our work constitutes the first analysis of the non-trivial effect of few-body perturbations in the 2d cluster state, which is of relevance for experimental proposals.   

\end{abstract}

\maketitle

{\it Introduction.-} What defines universal computationability of a quantum computer? Or in other words, when can a device be used to perform universal quantum compuation under practical assumptions? Certainly, this question is of paramount importance if we aim to develop large-scale quantum algorithms eventually. A way to address this problem is through the exciting approach of measurement-based 
quantum computation (MBQC) introduced by Raussendorf and Briegel \cite{mbqc}. In this setting, a quantum computation is performed by implementing local measurements on a highly-entangled quantum state, known as \emph{cluster state} \cite{briegel01}. As such, MBQC is a universal model of quantum computation. Therefore, it is relevant to know how much can one perturb the cluster state before it can no longer be used as a resource for quantum computation.  If we see the cluster state as the ground state of a many-body system, then the relevant question is whether the ground state of the perturbed system is still qualified for quantum computation or not. 

Unlike in 1d, the ground state of such a ``cluster" Hamiltonian in 2d (the 2d cluster state) is a universal resource for MBQC. The robustness of this state can be naturally analyzed at zero temperature by considering the effect of different Hamiltonian perturbations \cite{dohbar,son11,henning,else12}. The case of a magnetic field along the $x$ direction was first considered in Ref.~\cite{dohbar}, where the model was shown to be self-dual \cite{plenio}, and where the whole energy spectrum was itself also dual (up to degeneracies) to that of the quantum compass model \cite{Kugel82,Dorier05}, to the Xu-Moore model \cite{Xu04_05,Nussinov05_01}, and to the toric code in a transverse field \cite{vidal09b}. The phase transition in all these models takes place at the self-dual point and it is found to be first order \cite{aqocm,vidal09b}. The more intricate case of combined magnetic fields along $x$ and $z$ directions was later considered in Ref.~\cite{henning}. Here  a line of 1st-order transitions is detected in the $x-z$ plane ending at a quantum critical point. No sign of a phase transition along the $z$-axis or close to it is observed. 

Nevertheless, it still remains a mystery the way in which more general perturbations affect the quantum computational power of the 2d cluster phase. For instance, it is unknown whether the presence of few-body interactions modify significantly the phase diagram of the 2d system. This is also of practical importance, since experimental realizations of cluster states with quantum simulators may also be affected by the spurious presence of such few-body terms. 

Motivated by the above, we analyze in this letter the zero-temperature properties of the 2d cluster Hamiltonian on a square lattice under the effect of homogeneous Ising perturbations in the $z$ direction ($J_z$), and also under an homogeneous magnetic field in the $x$ direction ($h_x$). Unlike in the case of having magnetic fields only, we will see that introducing $J_z$ stabilizes a well-defined boundary for the cluster phase, which sets an upper bound for the amount of perturbation in the system so that the ground state is still useful for MBQC. More precisely, starting at $h_x = 0$ we find a line of 2nd order transitions in the 3d Ising universality class, which switches to a line of 1st order transitions at intermediate values of $h_x$ (tricritical end point), and continues until the self-dual point $(J_z = 0, h_x = J)$ where $J$ denotes the energy scale of the cluster Hamiltonian (see below). Our conclusions are further reinforced by calculations of the combined critical exponent $z\nu$. Moreover, we also compute the local fidelity \cite{zanardi06,zhou08a,zhou08b,klagges12} with the unperturbed 2d cluster state. Regarding the methods, our analysis proceeds in two steps. First, an approximate dual Hamiltonian describing the quasiparticle dynamics is derived which is exact for both limiting cases that only one of the perturbations is present. Most importantly, for $h_x=0$, the model reduces to the well-known transverse field Ising model establishing the existence of a 2nd-order phase transition. A classical approximation on this dual Hamiltonian is then done to gain intuition about the perturbed cluster Hamiltonian. Second, the phase diagram of the quantum model is computed by combining series expansions \cite{lo, taka} and variational infinite Projected Entangled-Pair States (iPEPS) methods \cite{iPEPS}, similar to the approach taken in Refs.~\cite{dusuel10,schulz12, henning}. Our work also constitutes the first analysis of the non-trivial effect of 2-body perturbations in the 2d cluster state. 

Let us stress that our work relies heavily on the methods and notations that we introduced in our previous work Ref.~\cite{henning}. Therefore, technical details (e.g. about our numerical methods) are avoided whenever possible in this manuscript. The interested reader is referred to Ref.~\cite{henning} for more details. 

{\it The model.-} We consider the cluster state Hamiltonian on an infinite square lattice under Ising and magnetic field perturbations. The cluster Hamiltonian can be written as a sum of mutually commuting stabilizers $K_{\mu} \equiv \sigma^x_{\mu} \bigotimes_{j \in \Gamma(\mu)} \sigma^z_j$, where $\Gamma(\mu)$ denotes the four nearest-neighbour spins of lattice site $\mu$ and the $\sigma^{\alpha}$ are the usual spin-$1/2$ Pauli matrices with $\alpha \in \{ x,y,z \}$. In terms of these stabilizers, the cluster Hamiltonian is given by $H_{\rm CL} = -J \sum_{\mu} K_{\mu}$. Since this is a five-body Hamiltonian, one may be tempted to say that its relevance is relative because it is hard to be produced in nature. This would be an incorrect conclusion, since this Hamiltonian is actually the low-energy effective theory of a ``more natural" Hamiltonian consisting only of two-body terms \cite{BartlettRudolph, GriffinBartlett}. Adding nearest-neighbour Ising interactions in $z$-direction and a magnetic field in $x$-direction, the Hamiltonian studied in this work reads
\beq
H = H_{\rm CL} - J_{z} \sum_{\langle i,j \rangle} \sigma^z_i \sigma^z_j - h_x \sum_i \sigma^x_i \ ,
\label{cl}
\eeq
with $J_z$ and $h_x$ being the Ising and magnetic field strengths, and where the sum in the Ising term runs over pairs of nearest-neighbours $\langle i,j \rangle$. Unless said otherwise, we will always consider the case $J=1$. 

At $h_x = J_z = 0$, the unique ground state of the above Hamiltonian is the 2d cluster state on a square lattice, which is the $+1$ eigenstate of all the stabilizer $K_{\mu}$ operators. For this case the model is exactly solvable, and excitations correspond to violations of the stabilizer constraints. The latter are hardcore bosons and are called ``clusterons" ($-1$ eigenvalues of some $K_{\mu}$ operators) \cite{henning}. However, a finite value of either $J_z$ or $h_x$ destroys the exact solvability of the model. At very large $h_x$ the system is clearly in a polarized phase along the $x$-direction, whereas at very large $J_z$ the system is in a symmetry-broken Ising phase. The nicely entangled structure of the 2d cluster state is therefore destroyed if these perturbations are large enough, in turn leading to a macroscopic rearrangement of the ground-state properties.  

Hamiltonian (\ref{cl}) is self-dual under the exchange of $J$ and $h_x$ for {\it any} value of $J_z$. This property is readily seen by the observation that the self-duality for $J_z=0$ is proven by performing a controlled-$Z$ operation on the Hamiltonian (\ref{cl}) \cite{henning}. Consequently, as the Ising term commutes with controlled-$Z$, the self-duality holds for any value $J_z$. It is therefore important to understand how these different limits are connected when both perturbations are present and to what extent the computationability of the perturbed cluster state is affected. 

In order to gain a first understanding of the phase diagram, we switch to an approximative description of the clusteron dynamics which contains the energetic properties of the two limiting cases, where only one perturbation is present, exactly. This is done by the use of pseudo-spins 1/2 $\tau^z_\mu$ on an effective square lattice which contains the eigenvalues $\pm 1$ of $K_\mu$ operators. The cluster Hamiltonian is therefore mapped to an effective field term. The action of the two perturbations $J_z$ and $h_x$ is as follows. The Ising coupling $J_z$ flips the two eigenvalues of nearest-neighbor $K_\mu$ operators. In contrast, the magnetic field $h_x$ on a site $i$ flips the four eigenvalues $K_j$ where $j$ is a nearest neighbor of $i$. In total, one obtains

\beq
H_{\rm eff} = -J \sum_i \tau^z_i - J_{z} \sum_{\langle i,j \rangle} \tau^x_i \tau^x_j - h_x \sum_{i,j_\alpha\in\Gamma (i)} \tau^x_{j_1}\tau^x_{j_2}\tau^x_{j_3}\tau^x_{j_4}  \ .
\label{eff}
\eeq         

For $h_x=0$, Eq.~(\ref{eff}) corresponds to a transverse field Ising model. In the other limit $J_z=0$ two decoupled Xu-Moore models are found. Let us stress that Hamiltonian (\ref{cl}) is isospectral to Eq.~(\ref{eff}) for these two cases which is not true for the general case because the sign of matrix elements not only depends on the clusteron configuration. 

From this self-duality important conclusions can be drawn. The perturbed cluster Hamiltonian for $h_x=0$ is isospectral to the transverse field Ising model on the square lattice. This implies a 2nd-order quantum phase transition between the cluster phase and the symmetry-broken Ising phase. The latter transition is therefore fundamentally different to the strong 1st-order phase transition between the cluster phase and the polarized phase at the self-dual point $J=h_x$ for $J_z=0$.  

We start by a classical approximation (CA) of Hamiltonian (\ref{eff}) which is well suited for a CA because each phase of the model has a classical analogue. This is different for the original problem (\ref{cl}). Here the maximally-entangled cluster state has no classical counterpart. In the dual language, the cluster phase corresponds to a polarized phase where all pseudospins point in $z$-direction. If only $h_x$ is finite, one has two degenerate classical ground states where all pseudo-spins point either in $x$- or $-x$-direction. Finally, there are 196 classical ground states for the case when only $h_x$ is finite. Indeed, one has two decoupled sublattices and on each sublattice 14 classical ground states exist (number of spins pointing in $-x$-direction should be even for each sublattice). 

Interestingly, the CA can be done with a single-site unit cell. This is a consequence of the fact that the energetic effect of the effective field term $\propto J$ on pseudo-spins pointing in $x$- or $-x$-direction is exactly the same. One therefore replaces $(\tau^x,\tau^y,\tau^z)$ by a classical vector of unit length $(\sin\theta \cos\phi , \sin\theta \sin\phi, \cos\theta)$. The classical energy per site $e_0^{\rm cl}$ is then given by 
\begin{equation} 
 e_0^{\rm cl} = -J\cos\theta-J_z (\sin\theta \cos\phi)^2 -h_x (\sin\theta \cos\phi)^4 \quad . 
\end{equation}
The angle $\phi$ is obviously zero. Minimizing the remaining expression, one finds a line of 2nd-order phase transition emerging from the pure Ising case $h_x=0$ which switches to a line of 1st-order phase transitions for intermediate values of $h_x/J_z$ ending at the limiting case of two decoupled Xu-Moore models for $J_z=0$ (see these results, together with the ones from the forthcoming analysis, in Fig.~(\ref{Fig3})). Most importantly, the CA captures both limiting cases qualitatively correct. Moreover, a similar behaviour of the phase boundary is observed for the full quantum problem (\ref{cl}) which we focus on in the following.    

To this end a number of observables for the Hamiltonian (\ref{cl}) are determined in the thermodynamic limit. In particular, we will consider the ground state energy per site $e_0 \equiv \lim_{N \rightarrow \infty} \langle H \rangle_0/N$ and the fidelity per lattice site with the cluster state $d \equiv  \lim_{N \rightarrow \infty} |\braket{\psi_{{\rm CS}}}{\psi_0}|^{2/N}$, where $\langle \ \cdot \ \rangle_0$ denotes the expectation value in the ground state $\ket{\psi_0}$, $N$ is the number of sites in the system, and $\ket{\psi_{{\rm CS}}}$ is the 2d cluster state on the infinite square lattice. We will also be interested in the behavior of the one-particle gap $\Delta \equiv (e_1 - e_0)$ -- where $e_1$ is the energy per site of the first excitation --, as well as the critical exponent $z\nu$ defined by $\Delta \sim (J_z-J_z^{{\rm crit}})^{z \nu}$ at fixed $h_x$ \footnote{Let us remind that the $z$ is the \emph{dynamical critical exponent} which specifies how the dynamical properties of the system become critical, and $\nu$ is the ciritcal exponent for the divergence of the \emph{correlation length}.}.  

\begin{figure}
\begin{centering}
\includegraphics[width=8.5cm]{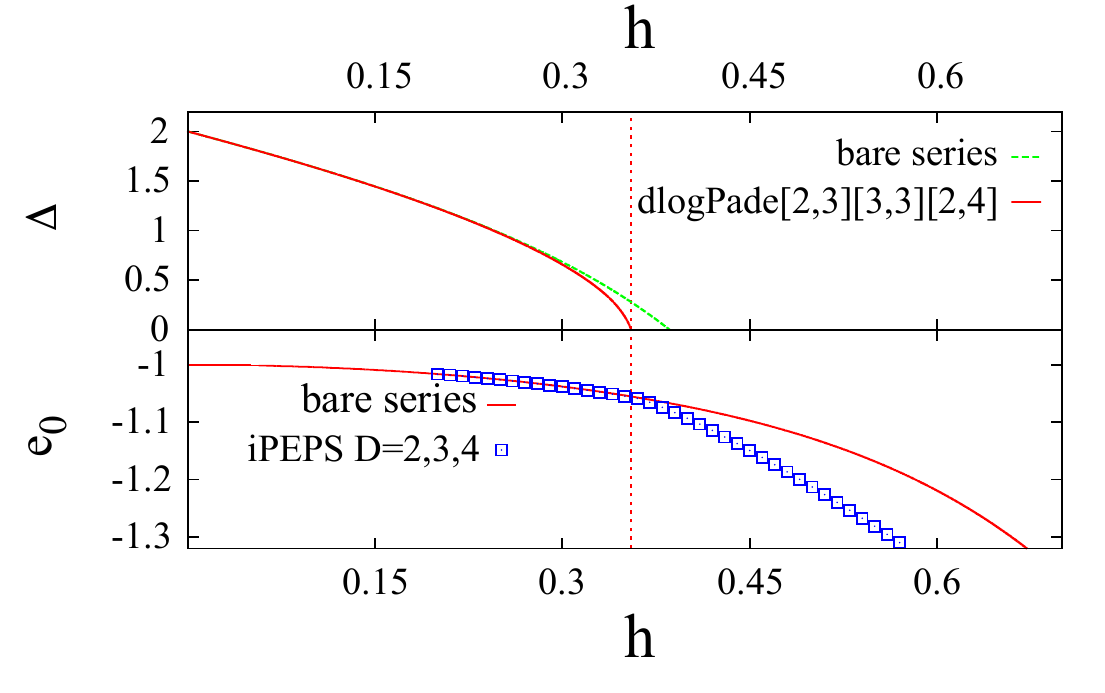}
\par\end{centering}
\caption{(Color online) Second-order phase transition for $J_z = h\sin(\alpha)$, $h_x = h\cos(\alpha)$, with $\alpha = 1.0$ and $J = 1$. We detect the transition at $h^{{\rm crit}} \approx 0.36$ (dashed line).}
\label{Fig1}
\end{figure}

{\it Series Expansions + iPEPS.-} Our approach to compute the phase diagram of the model in Eq.~(\ref{cl}) is a combined analysis of Series Expansions (SE) and iPEPS \cite{iPEPS}, which has already been successful in determining phase diagrams of a variety of systems \cite{dusuel10,schulz12, henning}. Following the same approach as in Ref.~\cite{henning}, we have computed a high-order series expansion of the ground state energy per site $e_0$, the one-particle gap $\Delta$, and the fidelity per site $d$. The series expansion is done about all three possible limits. Due to the self-duality, the large $h_x$ expansion can be obtained from the expansion inside the cluster phase for large $J$.  The energetic properties of the system have been computed using a partitioning technique provided by L\"owdin \cite{lo}, and the fidelity has been calculated using a projector method introduced by Takahashi \cite{taka}. From the behavior of the gap, we have also estimated the critical exponent $z\nu$ using the dlogPad\'e extrapolation. Explicitly, the dlogPad\'e extrapolation is based on the Pad\'{e} extrapolation of the logarithmic derivative of the one-particle gap $\Delta$
\begin{equation}
 \left[\frac{d}{dx}\ln \Delta \right]_{[L,M]}:=\frac{P_{L}}{Q_M}\quad ,
 \label{eq:dlog}
\end{equation}
where $P_{L}$ and $Q_M$ are polynomials of order $L$ and $M$ and $x$ denotes the perturbation parameter. Due to the derivative of the numerator in Eq.~(\ref{eq:dlog}) one requires $L+M=m-1$ where $m$ denotes the maximum perturbative order which has been calculated. The $\left[L,M\right]$ dlogPad\'e extrapolant is then given by
\begin{equation}
 {\rm dlog}\left[L,M\right] :=\exp\left(\int_0^x \frac{P_{L}(x')}{Q_M(x')} dx'\right)\quad .
 \label{eq:dlog2}
\end{equation}
In the case of a physical pole at $x_0$ one is able to determine the dominant power-law behaviour $|x-x_0|^{z\nu}$ close to $x_0$. The exponent $z\nu$ is then given by the residuum of $P_L/Q_M$ at $x=x_0$
\begin{equation}
 z\nu =\frac{P_{L}(x)}{\frac{d}{dx}Q_M(x)} |_{x=x_0}\quad .
 \label{eq:exp}
\end{equation} 

Regarding the iPEPS approach, we have employed the simplified update method in terms of Projected Entangled Pair Operators (PEPO) as described in detail in Ref.~\cite{henning}. This scheme allows us to implement easily the evolution in imaginary time driven by the 5-body, 2-body and 1-body terms in Eq.~(\ref{cl}). Also, expectation values have been extracted by using the directional Corner Transfer Matrix method from Ref.~\cite{dirctm}. Using this approach, we have computed $e_0$ and $d$ \cite{zanardi06,zhou08a,zhou08b,klagges12}. 

\begin{figure}
\begin{centering}
\includegraphics[width=8.5cm]{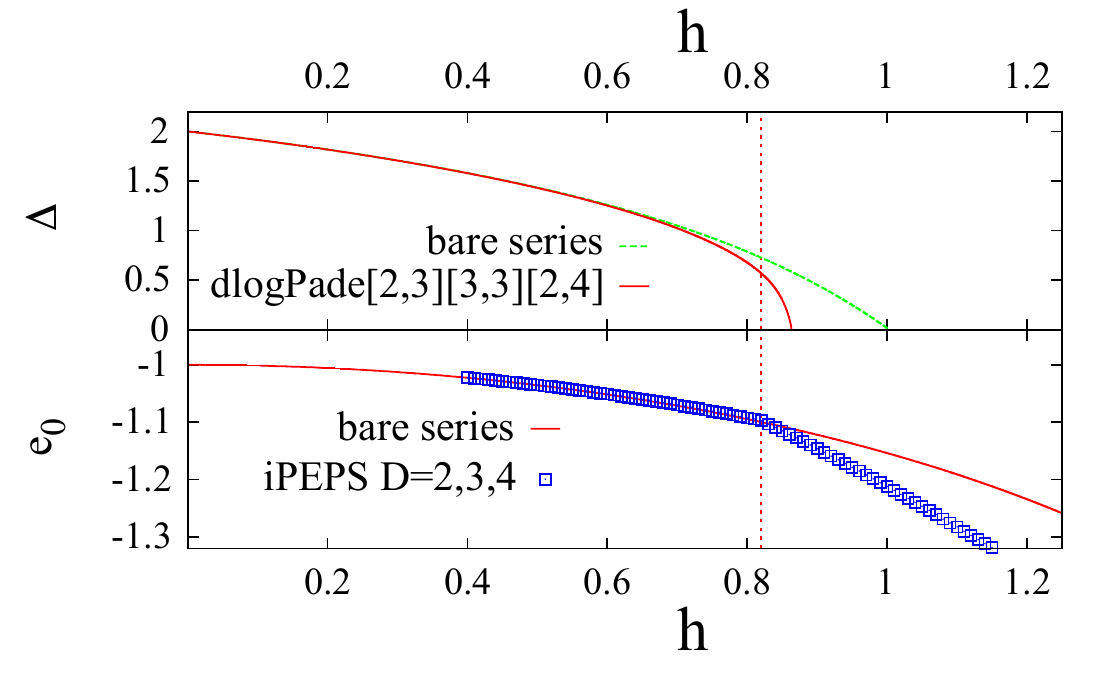}
\par\end{centering}
\caption{(Color online) First-order phase transition for $J_z = h\sin(\alpha)$, $h_x = h\cos(\alpha)$, with $\alpha = 0.2$ and $J = 1$.  We detect the transition at $h^* \approx 0.82$ (dashed line).}
\label{Fig2}
\end{figure}

\begin{figure}
\begin{centering}
\includegraphics[width=8.5cm]{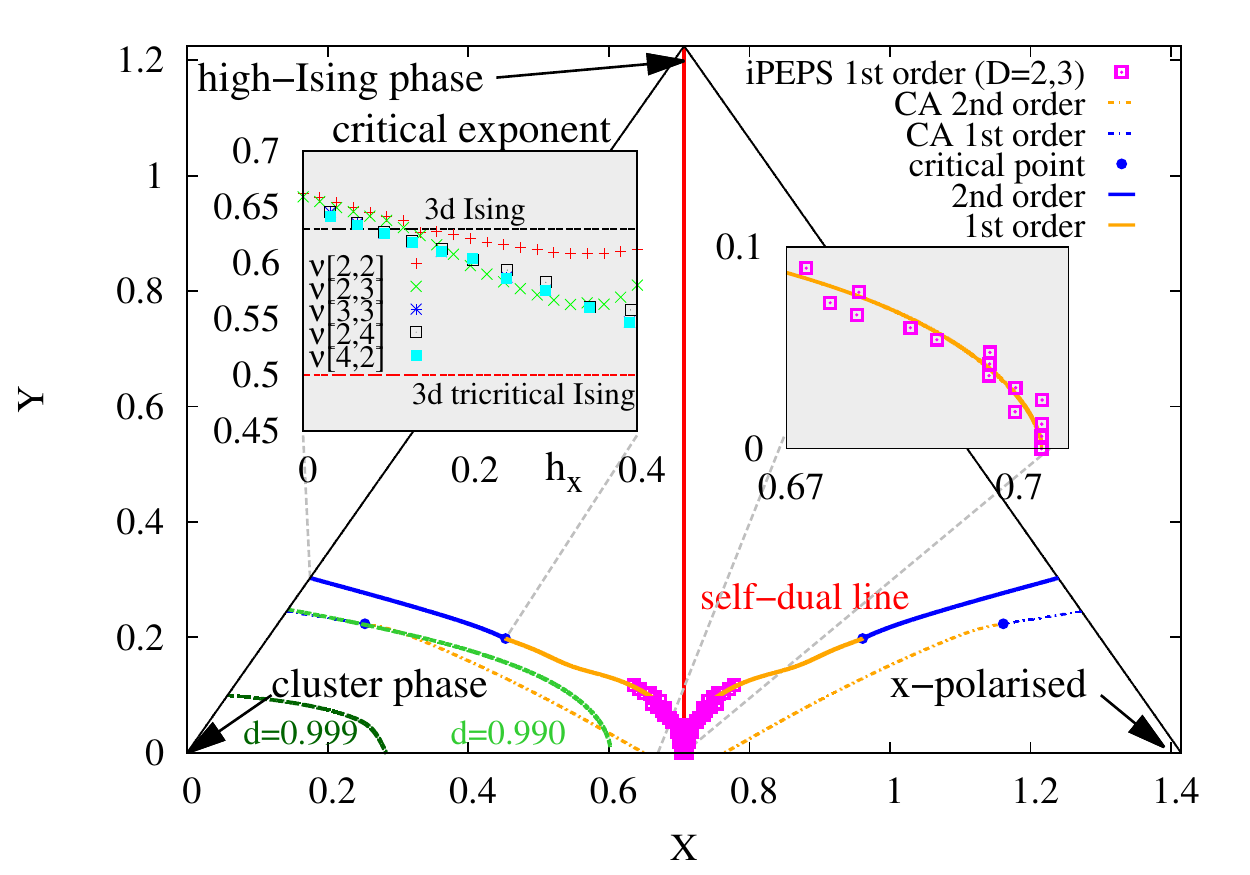}
\par\end{centering}
\caption{(Color online) Phase diagram as a function of variables $X = (1 - J_n + h_{x,n})/\sqrt{2}$ and  $Y = \sqrt{3/2}(1 - J_n - h_{x,n} )$ for $J = 1$ obtained by the combined Series Expansion plus iPEPS approach. The phase boundaries obtained within the CA are shown as dotted-dashed lines. Additionally, dashed lines refer to fidelities per site $d=0.999$ (black) and $d=0.99$ (green). Left inset shows the critical exponent $\nu$ for the divergence of the correlation length along the 2nd-order line in the phase diagram. The notation $\nu[p,q]$ indicates the type of resummation of the Series Expansion results in order to get different fittings to the exponent. Right inset is a zoom of the 1st-order transition points close to the self-dual line (vertical red line).}
\label{Fig3}
\end{figure}

The combined approach of series expansions plus iPEPS is done as follows. High-order series expansions of the one-particle gap $\Delta$ allows the location of 2nd-order quantum phase transition points. Setting $J_z = h \sin(\alpha), h_x = h \cos(\alpha)$ and fixing the angle $\alpha$, the critical perturbation value $h^{{\rm crit}}$ is defined by $\Delta(h^{{\rm crit}}) = 0$. This can be determined with good accuracy  by resummation techniques like dlogPad\'e extrapolations. However, series expansions restricted to one limit are not able to detect 1st-order phase transitions. We therefore define the field $h^*$ for which $e_0^{{\rm iPEPS}}(h) < e_0^{{\rm SE}}(h)$ with $h > h^*$ holds. The order of the phase transition is then assigned as follows: if $h^{{\rm crit}} < h^*$, then we conclude that there is a continuous phase transition at $h^{{\rm crit}}$ (see Fig.~(\ref{Fig1})). However, if $h^{{\rm crit}} > h^*$, then this indicates that the series expansion has missed a level crossing in the ground state observed in the variational energy computed by iPEPS, and therefore we conclude that there is a 1st-order phase transition at $h^*$ (see Fig.~(\ref{Fig2})). 

In practice, the series expansions in the cluster phase are done at order $7$ and are used for the comparison with iPEPS. Additionally, order $10$ (order $11$) has been obtained for the ground-state energy per site $e_0^{\rm Ising}$ (one-particle gap $\Delta^{\rm Ising}$) in the limit of large $J_z$ \cite{supp}. The iPEPS calculations is performed with bond dimension $2 \le D \le 4$ and with imaginary-time evolution with time steps of $\delta \tau = 10^{-4}$ \cite{iPEPS}. These approximations already provide sufficient accuracy for our purposes, with a typical relative error in the energy per site $e_0$ of $10^{-3} - 10^{-4}$. 
 
{\it Results.-} In Figs.~(\ref{Fig1})-(\ref{Fig2}) we show two representative examples of our calculations of $\Delta$ and $e_0$ for two directions in parameter space. The case of Fig.~(\ref{Fig1}) corresponds to $\alpha = 1.0$. Several dlogPad\'e's of the gap $\Delta$ are taken into account (upper panel) and one detects a critical point at $h^{{\rm crit}} \approx 0.36$ before $e_0^{{\rm iPEPS}}(h) < e_0^{{\rm SE}}(h)$, which is indicative of a continuous quantum phase transition at $h^{{\rm crit}}$. In Fig.(\ref{Fig2}), however, we observe a 1st-order transition. The calculation for this case is done for $\alpha = 0.2$, and we can see that the gap $\Delta$ closes well after $e_0^{{\rm iPEPS}}(h) < e_0^{{\rm SE}}(h)$, which happens at $h^* \approx 0.82$. 

The full phase diagram is presented in Fig.~(\ref{Fig3}). A coordinate transformation is used to map the 3d parameter space $(J, J_z, h_x)$ onto a 2d triangle which is given by $X= (1 - J_n + h_{x,n})/\sqrt{2}$, $Y = \sqrt{3/2}(1 - J_n - h_{x,n} )$ with the normalization $h_{x,n} + J_{z,n} + J_n = 1$ and also $J_n = J/(J + h_x + J_z)$ [$J_{z,n}$, $h_{x,n}$ accordingly]. Our calculations show that, starting from the Ising quantum critical point at $(J_z^{{\rm crit}} \approx 0.3285, h_x=0)$ \cite{isingcrit}, there is a critical line separating the cluster phase from a polarized phase. Along this line the critical exponent $z\nu$ is compatible with a value $\nu\approx 0.63$ (assuming $z=1$) indicating a 3d Ising universality class ending at a 3d Ising tricritical point having $\nu=1/2$ (see left inset). Here it is important to stress that series expansions typically slightly overestimate critical exponents. We also see in the phase diagram that at the critical point $( J_z \approx 0.27, h_x \approx 0.4 )$, the phase transition switches from 2nd to 1st order. The 1st-order phase boundary seems to end right at the self-dual point $(J_z=0 , h_x=1)$ (see right inset). This qualitative change from 2nd-order to 1st-order transition is also reflected in the behaviour of the gap $\Delta^{\rm Ising}$ in the Ising phase. Indeed, the corresponding momentum of the gap changes suddenly from $\vec{k}=(0,0)$ to $\vec{k}=(\pi,\pi )$ when increasing $h_x/J$. 

\begin{figure}
\begin{centering}
\includegraphics[width=8.5cm]{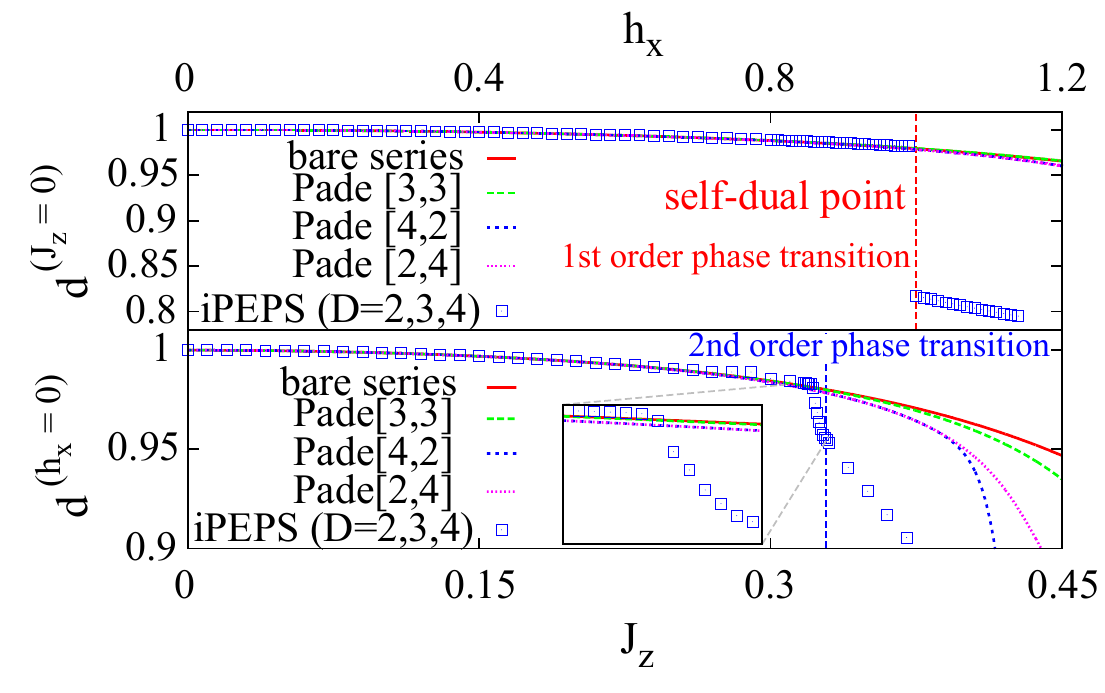}
\par\end{centering}
\caption{(Color online) Fidelity per site $d$ shown for the cases $J_z = 0$ (upper panel) and $h_x = 0$ (lower panel). The inset is a zoom of the region close to the 2nd order transition.}
\label{Fig4}
\end{figure}

Next, we show that this phase boundary may constitute a good upper bound for the usability of the perturbed cluster state as a resource state for MBQC. To this end the fidelity per site $d$ is shown for the limiting cases $J_z = 0$ (upper panel) and $h_x = 0$ (lower panel) in Fig.~(\ref{Fig4}). For $J_z = 0$ the iPEPS and SE data are in excellent agreement up to the self-dual point (dashed line). The fidelity per site $d$ is found to be quite large (e.g. $d \ge 0.99$) nearly within the whole cluster phase (see also Fig.~\ref{Fig3}). For $h_x = 0$ we find the fidelity rapidly decreasing down right at the 2nd-order phase transition. As before, the fidelity is quite large over most of the cluster phase.

{\it Conclusions.-} We have analyzed the quantum computational power of the perturbed 2d cluster state. This is done by accurately computing the phase diagram of the 2d cluster state Hamiltonian on a square lattice in the presence of an Ising interaction and a magnetic field. First, we derived an approximative model of the clusteron dynamics which is exact for both limiting cases. The cluster Hamiltonian plus Ising interactions is isospectral to the transverse field Ising model establishing a 2nd-order phase transition between the cluster and the symmetry-broken Ising phase. The full phase diagram of the quantum model is computed by series expansions and iPEPS. We have found very strong evidence of a well defined cluster phase, separated from polarized phases by a line of 1st- and 2nd-order transitions of 3d Ising type including a tricritical 3d Ising end point. This line of phase transitions sets an upper bound on the amount of perturbation that can be introduced in the system so that the ground state is still a valid resource for MBQC. Moreover, we also computed the local fidelity with the unperturbed 2d cluster state, showing that it is quite large almost through the entire cluster phase. The results in this work constitute the first analysis of the non-trivial effect of few-body perturbations in the 2d cluster state, which is of relevance for experimental proposals using for example ultracold atoms in optical lattices. Finally, let us suggest that it would be interesting to do a similar analysis with other states universal for MBQC \cite{gross07,wei11}.

\acknowledgements
RO acknowledges financial support from the EU through a Marie Curie International Incoming Fellowship. KPS acknowledges ESF and EuroHorcs for funding through his EURYI. Discussions with H. J. Briegel and A. L\"auchli are also acknowledged.

\newpage
\onecolumngrid
\section{Appendix}
{\it Series expansions.-} Let us briefly present the main series expansions in the limit $J \gg h_x,J_z$  (in units of $J = 1$). We obtained the groundstate energy per site
\begin{eqnarray}
e_0^{(7)} &=& -1-\frac{1}{8}\,{h_{{x}}}^{2}-\frac{1}{2}\,{J_{{{\it z}}}}^{2}-{\frac {13}{1536}}\,{h_{{x}}}^{4}-\frac{1}{32}\,{J_{{{\it z}}}}^{2}{h_{{x}}}^{2}-{\frac {15}{32}}\,{J_{{{\it z}}}}^{4}-{\frac {33}{64}}\,{J_{{{\it z}}}}^{4}h_{{x}}-{\frac {197}{98304}}\,{h_{{x}}}^{6}-{\frac {325}{36864}}\,{J_{{{\it z}}}}^{2}{h_{{x}}}^{4}\nonumber\\
&&-{\frac {3483}{8192}}\,{J_{{{\it z}}}}^{4}{h_{{x}}}^{2}-{\frac {147}{128}}\,{J_{{{\it z}}}}^{6}-{\frac {106535}{294912}}\,{J_{{{\it z}}}}^{4}{h_{{x}}}^{3}-{\frac {163}{64}}\,{J_{{{\it z}}}}^{6}h_{{x}}
\end{eqnarray}
and the gap up to order $7$:
\begin{eqnarray}
\Delta^{(7)} &=& 2-4\,J_{{{\it z}}}-\frac{1}{2}\,{h_{{x}}}^{2}-2\,{J_{{{\it z}}}}^{2}+\frac{3}{4}\,J_{{{\it z}}}{h_{{x}}}^{2}-\frac{9}{2}\,{J_{{{\it z}}}}^{2}h_{{x}}-3\,{J_{{{\it z}}}}^{3}-{\frac {15}{128}}\,{h_{{x}}}^{4}-{\frac {443}{128}}\,{J_{{{\it z}}}}^{2}{h_{{x}}}^{2}-{\frac {33}{8}}\,{J_{{{\it z}}}}^{3}h_{{x}}\nonumber\\
&&-\frac{9}{2}\,{J_{{{\it z}}}}^{4}+{\frac {323}{1152}}\,J_{{{\it z}}}{h_{{x}}}^{4}-{\frac {495}{256}}\,{J_{{{\it z}}}}^{2}{h_{{x}}}^{3}-{\frac {843}{512}}\,{J_{{{\it z}}}}^{3}{h_{{x}}}^{2}-{\frac {445}{32}}\,{J_{{{\it z}}}}^{4}h_{{x}}-11\,{J_{{{\it z}}}}^{5}-{\frac {575}{12288}}\,{h_{{x}}}^{6}\nonumber\\
&&-{\frac {197669}{110592}}\,{J_{{{\it z}}}}^{2}{h_{{x}}}^{4}-{\frac {3143}{1536}}\,{J_{{{\it z}}}}^{3}{h_{{x}}}^{3}-{\frac {72887}{3072}}\,{J_{{{\it z}}}}^{4}{h_{{x}}}^{2}-{\frac {14391}{512}}\,{J_{{{\it z}}}}^{5}h_{{x}}-{\frac {2625}{128}}\,{J_{{{\it z}}}}^{6}+{\frac {324187}{2654208}}\,J_{{{\it z}}}{h_{{x}}}^{6}\nonumber\\
&&-{\frac {1217425}{1327104}}\,{J_{{{\it z}}}}^{2}{h_{{x}}}^{5}-{\frac {5806163}{5308416}}\,{J_{{{\it z}}}}^{3}{h_{{x}}}^{4}-{\frac {11620739}{442368}}\,{J_{{{\it z}}}}^{4}{h_{{x}}}^{3}-{\frac {307057}{9216}}\,{J_{{{\it z}}}}^{5}{h_{{x}}}^{2}-{\frac {155251}{2048}}\,{J_{{{\it z}}}}^{6}h_{{x}}\nonumber\\
&&-{\frac {14771}{256}}\,{J_{{{\it z}}}}^{7}\quad\text{.}
\end{eqnarray}
Furthermore we computed the fidelity per site in order $6$ as
\begin{eqnarray}
d^{(6)} &=& 1-\frac{1}{8}\,{J_{{{\it z}}}}^{2}-{\frac {1}{64}}\,{h_{{x}}}^{2}-{\frac {93}{256}}\,{J_{{{\it z}}}}^{4}-{\frac {13}{512}}\,{J_{{{\it z}}}}^{2}{h_{{x}}}^{2}-{\frac {137}{36864}}\,{h_{{x}}}^{4}-{\frac {113}{256}}\,{J_{{{\it z}}}}^{4}h_{{x}}-{\frac {2961}{2048}}\,{J_{{{\it z}}}}^{6}-{\frac {271261}{589824}}\,{J_{{{\it z}}}}^{4}{h_{{x}}}^{2}\nonumber\\
&&-{\frac {6007}{589824}}\,{J_{{{\it z}}}}^{2}{h_{{x}}}^{4}-{\frac{20171}{14155776}}\,{h_{{x}}}^{6}\text{,}
\end{eqnarray}
and we also obtained the groundstate energy per site in the limit $J _z\gg h_x,J$  (in units of $J_z = 1$) in order 10:
\begin{eqnarray}
e_0^{(10)} &=& -1-5\,{h_{{x}}}^{2}-{\frac {5}{192}}\,{h_{{x}}}^{4}-{\frac {5}{12288}}\,{h_{{x}}}^{6}-{\frac {905}{28311552}}\,{h_{{x}}}^{8}-{\frac {1388129}{1630745395200}}\,{h_{{x}}}^{10}-10\,h_{{x}}J+{\frac {25}{16}}\,{h_{{x}}}^{3}J\nonumber\\
&&-{\frac {845}{18432}}\,{h_{{x}}}^{5}J-{\frac {2023}{1179648}}\,{h_{{x}}}^{7}J+{\frac {305}{96}}\,{h_{{x}}}^{2}{J}^{2}-{\frac {14155}{12288}}\,{h_{{x}}}^{4}{J}^{2}+{\frac {274619}{2359296}}\,{h_{{x}}}^{6}{J}^{2}+{\frac {25}{16}}\,h_{{x}}{J}^{3}-{\frac {20395}{9216}}\,{h_{{x}}}^{3}{J}^{3}\nonumber\\
&&+{\frac {1330063}{1179648}}\,{h_{{x}}}^{5}{J}^{3}-{\frac {14155}{12288}}\,{h_{{x}}}^{2}{J}^{4}+{\frac {28578437}{14155776}}\,{h_{{x}}}^{4}{J}^{4}-{\frac {845}{18432}}\,h_{{x}}{J}^{5}+{\frac {1330063}{1179648}}\,{h_{{x}}}^{3}{J}^{5}\nonumber\\
&&+{\frac {274619}{2359296}}\,{h_{{x}}}^{2}{J}^{6}-{\frac {2023}{1179648}}\,h_{{x}}{J}^{7}+{\frac {13495919}{163074539520}}\,{h_{{x}}}^{9}J-{\frac {31945903}{543581798400}}\,{h_{{x}}}^{8}{J}^{2}-{\frac {41771704639}{203843174400}}\,{h_{{x}}}^{7}{J}^{3}\nonumber\\
&&-{\frac {1035254507081}{815372697600}}\,{h_{{x}}}^{6}{J}^{4}-{\frac {32158658557}{15099494400}}\,{h_{{x}}}^{5}{J}^{5}-{\frac {1035254507081}{815372697600}}\,{h_{{x}}}^{4}{J}^{6}-{\frac {41771704639}{203843174400}}\,{h_{{x}}}^{3}{J}^{7}\nonumber\\
&&-{\frac {31945903}{543581798400}}\,{h_{{x}}}^{2}{J}^{8}+{\frac {13495919}{163074539520}}\,h_{{x}}{J}^{9}-5\,{J}^{2}-{\frac {5}{192}}\,{J}^{4}-{\frac {5}{12288}}\,{J}^{6}-{\frac {905}{28311552}}\,{J}^{8}\nonumber\\
&&-{\frac {1388129}{1630745395200}}\,{J}^{10}\quad\text{.}
\end{eqnarray}
In this limit we found a jump of the critical $k$-mode which corresponds to the gap. In the following we present the energies of the first excited state (which implies the existence of four quasi particles in the Ising-phase) for both $k$-modes $(0,0)^T$ and $(\pi,\pi)^T$ in order 8:
\begin{eqnarray}
\Delta^{\rm Ising}_{(0,0)} &=& 8-\frac{3}{4}\,{J}^{2}+\frac{13}{2}\,h_{{x}}J-\frac{3}{4}\,{h_{{x}}}^{2}+{\frac {43}{768}}\,{J}^{4}-{\frac {119}{64}}\,h_{{x}}{J}^{3}+{\frac {65}{384}}\,{h_{{x}}}^{2}{J}^{2}-{\frac {119}{64}}\,{h_{{x}}}^{3}J+{\frac {43}{768}}\,{h_{{x}}}^{4}\nonumber\\
&&-{\frac {19993}{884736}}\,{J}^{6}+{\frac {154661}{442368}}\,h_{{x}}{J}^{5}-{\frac {564215}{884736}}\,{h_{{x}}}^{2}{J}^{4}+{\frac {806611}{221184}}\,{h_{{x}}}^{3}{J}^{3}-{\frac {564215}{884736}}\,{h_{{x}}}^{4}{J}^{2}\nonumber\\
&&+{\frac {154661}{442368}}\,{h_{{x}}}^{5}J-{\frac {19993}{884736}}\,{h_{{x}}}^{6}+{\frac {82873487}{10192158720}}\,{J}^{8}-{\frac {153477341}{1274019840}}\,h_{{x}}{J}^{7}+{\frac {2091266777}{2548039680}}\,{h_{{x}}}^{2}{J}^{6}\nonumber\\
&&-{\frac {4309928299}{1274019840}}\,{h_{{x}}}^{3}{J}^{5}+{\frac {1884885289}{1019215872}}\,{h_{{x}}}^{4}{J}^{4}-{\frac {4309928299}{1274019840}}\,{h_{{x}}}^{5}{J}^{3}+{\frac {2091266777}{2548039680}}\,{h_{{x}}}^{6}{J}^{2}\nonumber\\
&&-{\frac {153477341}{1274019840}}\,{h_{{x}}}^{7}J+{\frac {82873487}{10192158720}}\,{h_{{x}}}^{8}\quad\text{,}
\end{eqnarray}
respectively
\begin{eqnarray}
\Delta^{\rm Ising}_{(\pi,\pi)} &=& 8+\frac{1}{4}\,{J}^{2}+\frac{1}{2}\,h_{{x}}J+\frac{1}{4}\,{h_{{x}}}^{2}-{\frac {5}{768}}\,{J}^{4}-{\frac {23}{64}}\,h_{{x}}{J}^{3}-{\frac {271}{384}}\,{h_{{x}}}^{2}{J}^{2}-{\frac {23}{64}}\,{h_{{x}}}^{3}J-{\frac {5}{768}}\,{h_{{x}}}^{4}+{\frac {5}{32768}}\,{J}^{6}\nonumber\\
&&+{\frac {5975}{147456}}\,h_{{x}}{J}^{5}+{\frac {14027}{32768}}\,{h_{{x}}}^{2}{J}^{4}+{\frac {57169}{73728}}\,{h_{{x}}}^{3}{J}^{3}+{\frac {14027}{32768}}\,{h_{{x}}}^{4}{J}^{2}+{\frac {5975}{147456}}\,{h_{{x}}}^{5}J+{\frac {5}{32768}}\,{h_{{x}}}^{6}\nonumber\\
&&-{\frac {17737}{1132462080}}\,{J}^{8}-{\frac {146069}{141557760}}\,h_{{x}}{J}^{7}-{\frac {22531727}{283115520}}\,{h_{{x}}}^{2}{J}^{6}-{\frac {73353331}{141557760}}\,{h_{{x}}}^{3}{J}^{5}-{\frac {33190037}{37748736}}\,{h_{{x}}}^{4}{J}^{4}\nonumber\\
&&-{\frac {73353331}{141557760}}\,{h_{{x}}}^{5}{J}^{3}-{\frac {22531727}{283115520}}\,{h_{{x}}}^{6}{J}^{2}-{\frac {146069}{141557760}}\,{h_{{x}}}^{7}J-{\frac {17737}{1132462080}}\,{h_{{x}}}^{8}\quad\text{.}
\end{eqnarray}
All series in this limit are clearly symmetric under the exchange $h_x \leftrightarrow J$ due to the self-duality of the model.
\end{document}